# OVERVIEW OF THE HEAVY ION FUSION PROGRAM[*]

C.M. Celata, Ernest Orlando Lawrence Berkeley National Laboratory, Berkeley, CA 94720, USA
for the U.S. Virtual National Laboratory for Heavy Ion Fusion

*Abstract*

The world Heavy Ion Fusion (HIF) Program for inertial fusion energy is looking toward the development and commissioning of several new experiments. Recent and planned upgrades of the facilities at GSI, in Russia, and in Japan greatly enhance the ability to study energy deposition in hot dense matter. Worldwide target design developments have focused on non-ignition targets for nearterm experiments and designs which, while lowering the energy required for ignition, tighten accelerator requirements. The U.S program is transitioning between scaled beam dynamics experiments and high current experiments with power-plant-driver-scale beams. Current effort is aimed at preparation for the next-step large facility, the Integrated Research Experiment (IRE)-- an induction linac accelerating multiple beams to a few hundred MeV, then focusing to deliver tens of kilojoules to a target. The goal is to study heavy ion energy deposition, and to test all of the components and physics needed for an engineering test of a power plant driver and target chamber. This paper will include an overview of the Heavy Ion Fusion program abroad and a more in-depth view of the progress and plans of the U.S. program.

## 1 INTRODUCTION

The international program in Heavy Ion Fusion is at the threshold of planning and constructing experiments which will test many accelerator and target issues in parameter ranges relevant to an eventual powerplant. In Europe, Russia, and Japan the emphasis is on measurements of stopping of heavy ions in matter. Accelerator studies there focus on the challenge of delivering very high-current beams to the target. In the U.S., small scaled experiments have been completed, and a series of accelerator experiments using driver-scale beams are in the construction and design phases. This paper will concentrate on the U.S. program, since it relies on a linac approach and has an accelerator physics program dedicated to energy production research. Note that much of the work cited here, and much more detail on HIF can be found in [1].

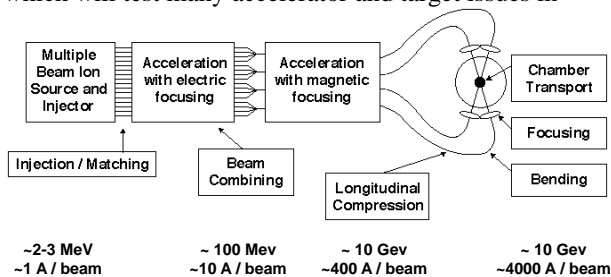

Figure 1: Induction linac driver subsystems & parameters

## 2 HIF RESEARCH IN THE U.S.

### 2.1 The Induction Linac Approach

Present Heavy Ion Fusion indirect drive targets require 1-7 MJ of heavy ions delivered to the target in about 10 ns, at a kinetic energy ~ 2-10 GeV. The total charge implied by this requirement leads to significantly higher line charge densities than can be stored in a single storage ring. While European designs have concentrated on combining and compressing pulses from several rings in the last phase of transport to the target, the U.S. approach is to make use of the efficiency of the induction linac for transporting high current beams in a single multibeam linac. A schematic view of one possible power plant accelerator (driver) is shown in Fig. 1. An injector provides multiple (~30-200) beams of heavy ions (e.g., $Cs^+$) at ~ 1.5-2 MeV, with current per beam of approximately 1 A. The beams are space-charge-dominated, with tune depressed by space charge to ~1/10 of the single particle tune. They are accelerated in parallel through induction cores which encircle the array of beams. Each beam is individually focused by quadrupoles-- in the example of Fig. 1, electrostatic quadrupoles at low energy, with a transition to magnetic quadrupoles at 50-100 MeV. Maximizing the transverse current density of the beam array, thereby minimizing the induction core radius, is important in controlling the cost of the accelerator. But electrostatic quadrupoles optimize at a smaller aperture than is optimal for high overall current density in magnetic quadrupoles. Therefore a 4-to-1 transverse combining of beams is included at the transition to magnetic focusing. After combining, the beam is accelerated to its final energy, then compressed longitudinally by a factor ~ 10 to obtain the short pulse required by the target. It is of utmost importance to keep the emittance growth in the accelerator low, in order that the beams can be focused to a spot of a few millimeter radius at the target. Desirable final normalized emittance is about 20 mm-mrad.

---

[*] This work supported by the Office of Energy Research, U.S. Department of Energy, under contract number DE-AC03-76SF00098.

## 2.2 Small-scale experiments

Earlier experiments [2,3] showed successful stable transport and acceleration of space-charge-dominated beams with tune depression in the desired range. Recent experiments have followed this mode, scaling dimensions and currents in order to reduce cost, but keeping physics parameters such as tune, tune depression, and ratio of beam radius to vacuum pipe radius in driver regimes.

This year, scaled experiments demonstrating transverse beam combining [4] and final focus [5] were completed. The Combiner Experiment merged four space-charge-dominated beams into a single transport channel. At the beginning of the merged-beam transport channel the beam-edge to beam-edge separation was ~4mm, as planned for a full-scale driver combiner. Two plasma oscillations downstream, the resulting (I=9mA, E=160keV) merged $Cs^+$ beam had a low normalized emittance (0.2 mm-mrad) and was still space-charge-dominated. Beam loss (<10%) was due in part to alignment imperfections and in part to halo formation in the merging process. The phase space evolution was in good agreement with particle-in-cell (PIC) simulations. This experiment will enable us to design full-scale merging systems with confidence.

The scaled final focus experiment used a 160 keV $Cs^+$ beam to model the ballistic focus of the 10 GeV $Bi^+$ beams in the HIBALL-II [6] driver design. Particle energy and mass, beam current, beam emittance, and focusing field were scaled to replicate the physics at one-tenth scale. The experiment successfully reproduces the calculated emittance-dominated focal spot size on target. A study was performed to determine the sensitivity of the focal spot area to deviations from the nominal design momentum. Sensitivity was actually less than calculated. Finally, the experimental current was increased to four times the scaled driver value. The space charge at the focus was then neutralized by 80% with electrons from a hot filament to recover an emittance-dominated spot.

## 2.3 Target Design

The past two years have seen significant progress in target design. By reducing the size of the hohlraum relative to the size of the capsule, the predicted energy required to ignite the capsule and achieve high gain has been reduced from 6 MJ to 3 MJ, according to 2-D Lasnex calculations [7]. This comes at the cost of reduced spot size-- final focal spot for the design is an ellipse with radii of 1.0 and 2.8 mm-- but gain increased (from ~70 to 130). More recently, work has focused on "hybrid" targets, which attempt to use shine shields to shield the capsule from direct illumination, while capturing the energy behind the shield for the hohlraum. In this design, the beams can have nearly the dimensions of the hohlraum, thus increasing the spot size over previous designs. Gain is lower (58) and energy higher (6.7 MJ), with spot radii of 3.8 and 5.4 mm. Work is also underway to smooth and simplify the required beam time profile.

## 2.4 Chamber Transport

The present fusion chamber design uses jets of liquid salt (Flibe, a mixture of LiF and $BeF_2$) to protect the outer chamber wall and final focus magnets from neutrons, and to serve as a heat transfer and tritium breeding medium [8]. Stripping of the beams as they pass through Flibe vapor, photoionization by x-rays from the target, possible plasma instabilities, and associated neutralization all must be investigated. Previous calculations [9] pointed to chamber vapor pressure ranges in which instabilities were absent. However the old calculations did not simultaneously include all of the phenomena listed above, and new high-gain close-coupled targets push toward lower beam kinetic energy, thus changing the parameter regime. A simulation program has begun, using the electromagnetic multispecies PIC codes LSP [10] and BPIC [11]. Both codes can be used for 2 or 3 dimensional simulations, and include electron dynamics, beam and plasma ions, and collisional ionization. Photoionization is being added.

Present accelerator designs have higher dimensionless perveance than previous designs, in part due to the lower kinetic energy at the target. As a result, neutralization is not only desirable, but also necessary, for the final transport. (Note, however, that driver designs with lower, but acceptable, target gain, and higher beam kinetic energy are still viable, but may have higher cost, if the lower energy beams prove difficult to focus.) Figure 2 gives results of an LSP simulation, showing near-ballistic focusing due to strong neutralization. Uniform Flibe vapor density of $10^{14}$ $cm^{-3}$ and Flibe plasma of density $10^{13}$ $cm^{-3}$ have been included in the simulation, and provide the electrons for neutralization. The beam is shown at four different times. The results, showing that most of the particles focus to within a 3 mm radius spot, are grounds for optimism, but it will be important to add photoionization and to obtain more accurate ionization cross sections. Separate calculations show that photoionization of gas near the target aids in neutralization, counteracting the effect of stripping. Excellent progress is being made in this area, with, so far, somewhat encouraging results.

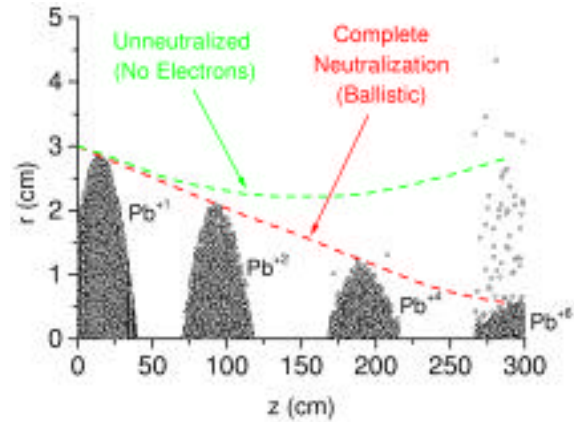

Figure 2: Simulation of Pb beam chamber transport.. Average charge state and beam profile shown at 4 points.

## 2.5 Technology Progress

Important to the ultimate success of the heavy ion fusion induction linac concept is the minimization of the cost of multibeam quadrupole arrays, insulators, ferromagnetic materials for induction cores, and pulsers. Electrostatic quadrupole arrays have been under development for quite a while, and satisfactory designs have been tested and used in experiment. Superconducting quadrupoles are presently at the design stage, with coils from a design by Advance Magnet Laboratory now being tested at MIT. A different design from LLNL will be tested next fiscal year. Insulators of about 0.5 m diameter from Advanced Ceramics have been fabricated and are awaiting testing. More pulser research is needed to reduce cost, but is presently not possible due to lack of funding.

Progress in ferromagnetic materials is reviewed in the paper by Caporaso in these proceedings. Three candidate materials have been identified. Measured efficiencies of 40-70% for 5 Hz 1 μs pulses are extremely encouraging.

Important work is also being done using water to study the flow of the liquids which protect the fusion chamber walls. Scaled experiments [12] have begun to measure the stability of the boundary of simulated Flibe jets.

## 2.6 Physics and Design Issues

Though crucial beam physics issues in the accelerator have been tested in small-scale experiments, there remain important phenomena which need to be explored in order to have confidence in the induction linac approach to a heavy ion fusion driver. Issues that can not be examined at small scale and low kinetic energy include beam loading of the induction modules and associated longitudinal stability, and halo studies. Simulation shows that the resistive wall effect of the interaction of beams with induction modules gives slow growth of longitudinal waves and emittance, but experimental results are needed to confirm this and to determine the specifications for the feedforward system for stabilization. The effect of electrons from charge exchange with neutral gas and impact of halo with the walls is also an active area of study presently. Previous experiments used electrostatic focusing, where quadrupole fields clear the aperture of electrons. Electron physics must now be explored experimentally at higher velocities, where magnetic focusing is used. And the dynamic aperture for both focusing methods must be defined.

A further issue is the choice of ion source and injector design. A 2 MeV, 800 mA injector has been built, which produces a driver-scale beam. Though emittance is acceptable, beam hollowing has been measured, and an active program to improve the optics is underway. When this is completed, the 2 MeV injector will provide a proof-of-principle example for a driver injector. However the design must be scaled to produce a multibeam array. A competing design, which seeks to reduce the dimensions and cost of the injector, is being researched. This concept transversely merges ~100 small beams at ~1 MeV to form each beam, thus using the fact, known from the Child-Langmuir equation, that small-diameter sources will provide extremely bright beams.

For both of these injector concepts, various ion sources have been proposed. The 2 MeV injector uses an aluminosilicate source impregnated with cesium. This gives acceptable beam quality, but to attain acceptable lifetime a continuous feed of a solution containing the ion should be developed. Gas sources are also a possibility, and are being worked on. Issues here include risetime, which ideally should be <1 μs, and neutral emission, which could contribute to beam loss and optics changes due to charge exchange with the heavy ions.

Finally, many physics questions remain in the final focus and transport of the beam through the fusion chamber to the target. As indicated above, simulations are presently making great progress in this area. Experiments are planned and will be discussed below.

## 2.7 Future Plans

A program plan has been created to address the issues outlined above, and to demonstrate all the systems involved in a heavy ion fusion accelerator and fusion chamber. The approach involves a sequence of accelerator experiments performed in parallel with technology development and verification of target physics.

The first of the accelerator experiments, the High Current Experiment (HCX), is under construction at LBNL. This will be a 1-beam transport experiment using a coasting beam with current and pulse length similar to a driver beam (~500-800 mA, 1.5 MeV, ~10 μs). In Phase 1 the beam will propagate through ~20 lattice periods of electrostatic quadrupole transport, transitioning to a few periods of magnetic focusing. In Phase 2, ~100 magnetic quadrupoles will be added. The purpose of the experiment is to explore dynamic aperture in the two focusing systems, and in Phase 2, the role of electrons in magnetic transport. Limits on pulse length from the injector due to gas generated by halo impacting the wall will also be explored. The Phase 2 quadrupoles will most probably be superconducting. In order to quickly attain the velocity appropriate to magnetic focusing, a medium mass ion, $K^+$, will be used.

Succeeding the HCX will be a proposal for the Integrated Reseach Experiment (IRE). Together with target physics experiments, the IRE will test all of the systems necessary for an Engineering Test Facility (ETF). The ETF would be a prototype powerplant driver and chamber experiment.

As presently envisioned, the IRE will accelerate ~30-80 beams of $K^+$ to energies of 200-400 MeV. Beams will be driver-scale, i.e., ~ 1 A at ~2 Mev coming from the injector, increasing to ~600 A at the end of the machine. This will be the first experiment to reach the total currents and length necessary to look at beam loading and longitudinal stability, though the low growth rate for the longitudinal instability predicted by simulation makes it likely that this instability will be studied by deliberately enhancing the growth rate by introducing longitudinal perturbations. The IRE will also explore neutralization after the final focus, and perhaps plasma channel transport. An important goal of this machine is to address target issues that are particular to heavy ions. Experiments will investigate the beam-plasma interaction, and if possible, depending on funding/machine capability, issues such as fluid instabilities for direct drive. Exploring the final focus and transport requires perveance per beam of at least $10^{-4}$, and the target experiment goals lead to intensity at the target of $> 3 \times 10^{12}$ W/cm$^2$ and total energy of several kilojoules.

Two projects are presently actively preparing for the IRE design: theoretical activity to provide end-to-end 3-D simulation capability for this machine, and an injector-development project which is simulating, and soon will be experimentally investigating, the two injector designs mentioned above. A multibeam injector module will be built and tested by about 2003. Simulation thus far has explored alignment tolerances and various magnet nonlinearities, all of which seem to produce very little beam degradation. It is clear, however, that beam steering will be needed in the magnetic-focused section.

In addition to the program at LBNL, a scaled experiment is being constructed at University of Maryland [13] which uses an electron ring to model long length scale propagation of high perveance beams.

In parallel with the accelerator development described above, target development will proceed, based on simulation and benchmarked by results from the National Ignition Facility and the Omega facility at Rochester, and also by the high density beam-plasma experiments in Europe which will be described below.

## 3  THE HIF PROGRAM ABROAD

### 3.1 Introduction

Effort abroad in HIF is largely based on dual use of existing or planned nuclear physics facilities. Beams from these facilities are used primarily for experiments to measure the interaction of heavy ions with plasmas and dense hot matter. These experiments will improve the accuracy of calculations of both stripping in the fusion chamber and interaction of beams with the target. Accelerator design is also an active area of research. The major difference between the approach to the accelerator abroad and the approach in the U.S. has been alluded to above-- existing European and Japanese facilities consist of synchrotron and storage rings based on rf acceleration, and the program, "piggybacking" on those facilities, follows this approach. Therefore the main issue becomes pushing the space charge limits of the system in order to transport the required large currents to the target.

### 3.2 Russia

The Russian program [14] is preparing a major upgrade of the ITEP accelerator complex for acceleration of high current beams. The TeraWatt Accumulator project (ITEP-TWAC) will use a new laser ion source to produce about $5 \times 10^{10}$ ions per pulse. These are then accelerated in the I-3 pre-injector to 1.6 MV/u and injected into the 13 T-m UK booster ring. At 0.7 GeV/u the bunch is injected in single turn mode into the U-10 synchrotron. Non-Liouvillian injection using a foil stripper is employed to minimize phase space, and a kicker system is used to direct the beam onto the stripper only during injection. The bunch will then be compressed from 1 μs to ~ 100 ns and focused to a spot calculated to be ~ 1 mm in diameter. Precommissioning with demonstration of multi-turn stacking for $C^{6+}$ is scheduled for December of 2000.

A multi-dimensional 3-temperature hydrodynamic Eulerian code has been used to explore the capabilities of the TWAC facility for beam-plasma interaction studies relevant to HIF targets. Using a hollow beam on a hollow cylindrical target, calculations indicate the possibility of attaining high enough temperatures to produce measurable quantities of fusion neutrons.

Ion energy loss studies are presently underway in a collaboration at GSI, using C, Kr, and Pb beams from the UNILAC [15]. In the near future an evolution of these experiments to the densities produced by explosively-driven targets will take place. This technique has been tested with the 3 MeV protom beam at ITEP.

### 3.3 Europe

The European effort has been recently reviewed by I. Hofmann [16]. Much of HIF accelerator effort in the European community is centered at GSI Darmstadt where, as described in many papers in these proceedings, a high intensity upgrade is in progress which is based on a new IH-structure injector for 15 mA of $U^{4+}$. This upgrade, which will eventually lead to beams of 1 kJ $U^{28+}$ at 200 MeV/u, provides for beam-plasma experiments with 1-10 eV temperatures. A specially tailored plasma lens which can create hollow cylinder shaped ion beams will be used for cylindrical targets. Also of great interest for the beam compression needed at the end of a HIF driver is a prototype RF pulse compression cavity designed to deliver a 200 kV amplitude field at 0.8 MHz [17].

Studies are underway for the next-generation GSI accelerator complex, which has as its main focus the acceleration of radioactive beams and secondary beams (antiprotons), and production of medium-energy nuclear collisions. A proposed high-current accumulation and compression ring (HAR) for this facility is of interest for exploring accelerator issues associated with the handling of pulses of hundreds of amperes of heavy ions, as well as for its potential for beam-plasma experiments. As presently envisioned, this ring could produce several tens of kilojoules of $Xe^{54+}$, with pulse length after compression of 30-50 ns. This beam is projected to produce 30-50 eV temperatures in matter, greatly extending the parameter range of study for ion energy loss.

Along with these accelerator efforts, experiments are making progress with careful measurements of ion stopping, including charge exchange cross sections and dE/dx in regimes not previously accessible [18].

Theoretical results of note for HIF as well as other high-current applications are the studies by Boine-Frankenheim of GSI of nonlinear evolution of resistive impedance-driven longitudinal instability in circular machines [19]. Though more work remains to be done to cover the parameter space, Vlasov simulations show saturation of the instability with minimal emittance increase, for suficiently small resistive impedances.

Channel transport for final transport of a high current beam to the target is also being investigated at GSI, where a 13.6 MeV/u carbon beam from the UNILAC is transported through 50 cm of 10-20 mbar of ammonia gas [20].

European target design work continues, using both multi-radiator and 2-sided illumination hohlraums [21]. Variation of gain with pulse shape has been explored. Other results show that the high gain targets studied have high sensitivity to beam pointing errors.

*3.4 Japan*

The Japanese program centers on the use of existing and planned facilities to look at beam-plasma studies of ion stopping [22]. Experiments are conducted at TIT, RIKEN, and HIMAC, where projectile energies of 0.1 - 6 MeV/u are available. The future RIKEN/MUSES multi-use accelerator system, with the main goal of radioactive beam experiments, will be able to greatly increase the intensity of heavy ions available. Beams from this facility (e.g., $2.6 \times 10^{11}$ $_{238}U^{49+}$ ions with normalized emittance after cooling of 10 mm-mrad) should produce plasma temperatures of 10-50 eV.

Accelerator research of relevance to HIF in Japan includes work on laser ion sources, induction module waveform synthesis, and characterization and development of induction cores from the amorphous metal "FINEMET".